\documentclass[aps,pra,reprint,superscriptaddress]{revtex4-1}
\usepackage{amsmath,amsfonts,amssymb}
\usepackage{graphicx,calc,bm}
\usepackage[breaklinks,colorlinks,urlcolor=blue,citecolor=blue,linkcolor=blue]{hyperref}
\begin{document}
\title{Multiorder topological superfluid phase transitions in a two-dimensional optical superlattice}

\author{Yu-Biao Wu}
\author{Guang-Can Guo}
\affiliation{Key Laboratory of Quantum Information, and CAS Center for Excellence in Quantum Information and Quantum Physics, University of Science and Technology of China, Hefei, Anhui 230026, People's Republic of China}

\author{Zhen Zheng}\email{zhenzhen.dr@outlook.com}
\affiliation{Department of Physics and HKU-UCAS Joint Institute for Theoretical and Computational Physics at Hong Kong, The University of Hong Kong, Pokfulam Road, Hong Kong, China}

\author{Xu-Bo Zou}\email{xbz@ustc.edu.cn}
\affiliation{Key Laboratory of Quantum Information, and CAS Center for Excellence in Quantum Information and Quantum Physics, University of Science and Technology of China, Hefei, Anhui 230026, People's Republic of China}
\begin{abstract}

Higher-order topological superfluids have gapped bulk and symmetry-protected Majorana zero modes with various localizations.
Motivated by recent advances, we present a proposal for synthesizing multi-order topological superfluids that support various Majorana zero modes in ultracold atomic gases.
For this purpose,
we use the two-dimensional optical superlattice that introduces a spatial modulation to the spin-orbit coupling in one direction, providing an extra degree of freedom for the emergent higher-order topological state.
We find the topologically trivial superfluids, first-order and second-order topological superfluids, as well as
different topological phase transitions among them with respect to the experimentally tunable parameters.
Besides the conventional transition characterized by the Chern number associated with the bulk gap closing and reopening,
we find the system can support the topological superfluids with Majorana corner modes,
but the topological phase transition undergoes no gap-closing of bulk bands.
Instead, the transition is refined by the quadrupole moment and signaled out by the gap-closing of edge states.
The proposal is based on the $s$-wave interaction and is valid using existing experimental techniques,
which unifies multi-order topological phase transitions in a simple but realistic system.

\end{abstract}
\maketitle

\section{Introduction}

Ultracold atoms in optical lattices offer a remarkable platform for
investigating quantum many-body problems.
Based on existing optical techniques, the high controllability and lack of disorder in ultracold atomic gases make it an ideal platform for the discovery of unconventional topological phases that are difficult to be realized or evinced in ordinary condensed-matter systems \cite{Bloch2012nphys,Gross2017sci,Zhang2018advphys}.
In particular, the engineering of the artificial gauge fields \cite{Goldman2014rpp,Eckardt2017rmp} and nontrivial atomic interaction or pairing \cite{Regal2003prl,Kuns2011pra,Kraus2013prl,Liu2015prl,Zhang2015prl,Cui2017prl,Yao2019nphys} facilitates the investigations in a variety of phenomena and phases, including the topological Hall effect \cite{Palmer2008pra,Aidelsburger2011prl,Celi2014prl,Stuhl2015sci,Grass2015pra,Kollath2016prl}, topological insulators \cite{Stanescu2009pra,Kennedy2013prl}, topological semimetals \cite{Sun2011nphys,Dubcek2015prl}, and topological superfluids \cite{Gong2011prl,Qu2013ncomms,Zhang2013ncomms,Chan2017prl}.
Among them, the engineering topological superfluid has been attracted intensive interests
for its unconventional non-Abelian exchange statistics and the potential application in fault-tolerant topological quantum computation \cite{Nayak2008rmp}.

It has been well known that traditional $d$-dimensional topological superfluids support a bulk gapped phases with topologically-protected Majorana edge modes (MEMs) on the ($d-1$)-dimensional boundaries \cite{hasan2010colloquium},
which is specified as a first-order topological superfluid (FOTSF) phase.
In stark contrast,
for higher-order topological phases
\cite{schindler2018higher,zhang2013surface,benalcazar2014classification,benalcazar2017quantized,benalcazar2017electric,langbehn2017reflection,peng2017boundary,Slager2015}, both the $d$-dimensional bulk and the ($d-1$)-dimensional boundary are gapped, but the Majorana zero-energy modes (MZMs) arise at lower dimensions.
Particularly, the higher-order topological phases \cite{khalaf2019boundaryobstructed}, support MZMs whose distribution is localized at corners of the two-dimensional (2D) lattice, i.e., the Majorana corner modes (MCMs) instead of MEMs.
Such a topological phase is characterized by the second-order topological invariant, which is also known as the second-order topological superfluids (SOTSF).
It has recently attracted a great deal of attention because of the enrichment of boundary physics,
and a set of schemes based on solid-state systems as well as ultracold atomic gases has been proposed for realizing MCMs \cite{yan2018majorana,liu2018majorana,ezawa2018topological,zhu2018tunable,zhu2019second,ezawa2018magnetic,volpez2019second,wu2019plane,Kheirkhah2020prl,Wu2020pra,Wu2020prx,tiwari2020chiral,Wu2020jpcm}.
Most of those schemes rely on nontrivial atomic interaction (e.g., the higher-order-wave ones), which generally faces unexpected frustrations, such as inelastic three-body scattering \cite{Levinsen2008pra}, to be implemented with current experimental technologies.

The paper is organized as follows:
In Sec.\ref{sec-model}, we present the model Hamiltonian of the proposal in the SO coupled Fermi gas.
The engineering of homogeneous SO coupling has been experimentally realized \cite{Lin2011soc,Kolkowitz2016soc,Huang2016soc,Wu2016soc,Hamner2014soc}.
We extend these results and propose a feasible scheme
by using an A-B sublattice structure to impose a staggered modulation,
which is key for realizing the SOTSF phase.
In Sec.\ref{sec-res}, we show the phase diagrams and associated numerical results.
Assisted by the controllable staggered modulation and the tunable interaction strength,
we find the first- and second-order topological transitions governed by different bulk-boundary correspondences.
Especially during the second-order transition,
the SOTSF phase is distinguished not by gap-closings of the bulk bands but by those of the edge states.
In Sec.\ref{sec-dis},
we discuss the potential ways for engineering the model using existing methods in ultracold atoms,
and summarize the work.

\section{Lattice Model} \label{sec-model}

\begin{figure}[t]
	\centering
	\includegraphics[width=0.49\textwidth]{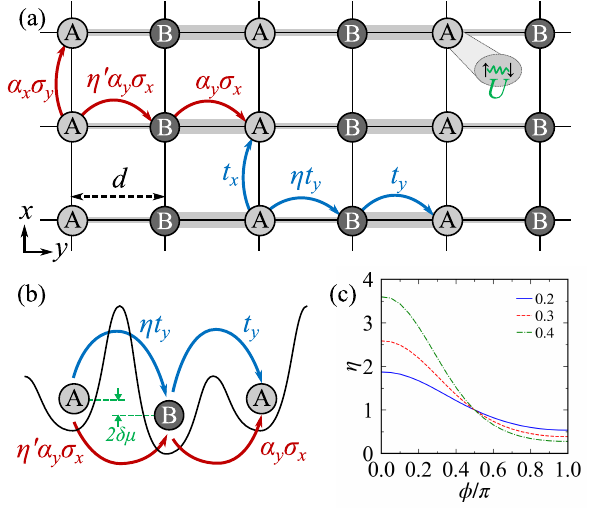}
	\caption{(a) Illustration of the lattice model.
	The system is dimerized in A-B sublattices in the presence of the double-well structure.
	(b) Detailed setup for NN interplays along the $y$ direction.
	Both the hopping and SO coupling $\alpha_y$ exhibit the staggered pattern.
	The energy offset between atoms in adjacent sites is characterized by $\delta\mu$.
	(c) The dimensionless parameter $\eta$ as a function of the phase $\phi$ for different $V_L'/V_L$.
	We set $V_L=4E_R$ with $E_R=\hbar^2k_L^2/(2m)$ being the recoil energy of the lattice \cite{Walters2013pra}.}
	\label{fig-model}
\end{figure}

We consider the spinful Fermi gas trapped in a two-dimensional (2D) optical lattice.
For such a fermionic system with contact interaction,
it has been known that by simultaneously introducing the SO coupling and the Zeeman field,
the $s$-wave interaction will become an effective chiral $p$-wave form if only the lower band is fully filled \cite{zhang2008p}.
When increasing the Zeeman field, the bulk band gap closes and reopens.
Thus the system evolves to FOTSF that supports MEMs \cite{Gong2011prl},
which motivates us to search the possible multi-order topological transitions based on such a FOTSF transition.
In comparison with previous works (see, e.g., Ref.\cite{zhang2013surface}), one can directly conclude that the above-mentioned system does not support the interesting SOTSF phase because of insufficient degrees of freedom.
However in general, an optical lattice can be dimerized in an A-B sublattice system via external optical fields \cite{Kennedy2013prl,Dubcek2015prl} or the manipulation of the lattice geometry \cite{Tarruell2012superlattice,Jotzu2014superlattice,Gorg2018superlattice}.
In this way, it provides extra degrees of freedom that are presented by the sublattice index.
Besides, the intersublattice interplays can be designed by the optical lattice configuration or laser fields.
It paves the way to engineer interesting Hamiltonians in ultracold atoms.
In this work, we follow this trick.
We prepare the trap potential as ($k_L=\pi/d$, $d$ is the lattice constant)
\begin{equation}
	V_L({\bf r})= V_L\sin^2(k_Lx) + V_L\sin^2(k_Ly) + V_L'\sin(k_Ly+\phi)
\end{equation}
Here $\phi$ is the relative phase between the two potentials.
With the setup of $V_L'<V_L$,
the perturbative potential $V_L'$ will not phonemically change the lattice configuration constructed by $V_L$.
Instead, in the presence of $V_L'$, the lattice model exhibits a double-well structure within adjacent sites along the $y$ direction.
It comprises an A-B sublattice configuration, which is illustrated in FIG.\ref{fig-model}.

The Fermi gases can be described by the tight-binding model Hamiltonian composed of three parts,
\begin{equation}
	H = H_0 + H_{\rm ext} + H_{\rm int} \,. \label{eq-h-site}
\end{equation}
The first part describes the nearest-neighbor (NN) hopping and chemical potential,
\begin{align}
	& H_0 = - \sum_{j,s}\mu \psi_{js}^\dag \psi_{js}
	+ \delta\mu (\psi_{2{\bf j}-\hat{\bf e}_y,s}^\dag \psi_{2{\bf j}-\hat{\bf e}_y,s}
	-\psi_{2{\bf j},s}^\dag \psi_{2{\bf j},s}) \notag\\
	& - (t_x \psi_{{\bf j},s}^\dag \psi_{{\bf j}+\hat{\bf e}_x,s}
	+ \eta t_y \psi_{2{\bf j}-\hat{\bf e}_y,s}^\dag \psi_{2{\bf j},s}
	+ t_y \psi_{2{\bf j},s}^\dag \psi_{2{\bf j}+\hat{\bf e}_y,s} +H.c.)
	\label{eq-h-0-site}
\end{align}
where $\psi_s$ denotes the field operator for fermionic atoms of pseudospin $s=\uparrow,\downarrow$.
$t_{x,y}$ is the hopping magnitude and is chosen as the energy unit.
$\hat{\bf e}_{x,y}$ stands for the unit vector.
$H.c.$ stands for the Hermitian conjugate.
$\mu$ is the chemical potential.
In Hamiltonian (\ref{eq-h-0-site}),
we introduce the dimensionless parameter $\eta$ and the energy offset $\delta\mu$ to characterize the spatial modulation of the hopping magnitude and chemical potential
due to the double-well structure.
We note that $\eta$ can be artificially tunable via not only the trap depths $V_L$ and $V_L'$, but also the relative phase $\phi$,
as shown in FIG.\ref{fig-model}(c).

The second part of Hamiltonian (\ref{eq-h-site}) describes the external fields including the Rashba-type SO coupling and the Zeeman field.
Its form is given by
\begin{align}
	&H_{\rm ext} = \sum_{j,s,s'} V_z\psi_{{\bf j}s}^\dag \sigma_z^{ss'} \psi_{{\bf j}s'} - (i\alpha_x \psi^\dag_{{\bf j},s}\sigma_y^{ss'}\psi_{{\bf j}+\hat{\bf e}_x,s'} +H.c.) \notag\\
	&+ [i\alpha_{y} (\eta' \psi^\dag_{2{\bf j}-\hat{\bf e}_y,s}\sigma_x^{ss'}\psi_{2{\bf j},s'}
	+\psi^\dag_{2{\bf j},s}\sigma_x^{ss'}\psi_{2{\bf j}+\hat{\bf e}_y,s'}) +H.c.]
\end{align}
where $\hat{\psi}\equiv ( \psi_\uparrow, \psi_\downarrow)^T$.
$\sigma_{x,y,z}$ are the Pauli matrices defined in the spin space.
$V_z$ is the strength of the Zeeman field.
$\alpha_{x,y}$ denotes the strength of the SO coupling.
Likewise, due to the double-well structure,
the magnitude of the SO coupling also exhibits a staggered pattern that is controllable via a dimensionless parameter $\eta'$.
For simplicity and without loss of generality,
hereinafter we shall choose $t_x=t_y=t$, $\alpha_x=\alpha_y=\alpha$, and set $\eta'\approx\eta$ to facilitate the further discussion.

The last part of the Hamiltonian (\ref{eq-h-site}) describes the $s$-wave interaction,
which is usually controlled by Feshbach resonances,
\begin{equation}
	H_{\rm int} = -\sum_{j} U \psi_{j\uparrow}^\dag \psi_{j\downarrow}^\dag \psi_{j\downarrow} \psi_{j\uparrow} \,. \label{eq-h-int-site}
\end{equation}
Here $U$ denotes the interaction strength.
The minus sign makes the interaction that we focus on attractive.
In cold atoms, this gives rise to the superfluid state, in which atoms form Cooper pairs
in the same way as electrons do in superconductors of solids.

\section{Numerical results} \label{sec-res}

To investigate nontrivial topological properties of the system, we employ the mean-field Bogoliubov-de Gennes (BdG) approach.
By introducing the order parameters $\Delta_j=-U\langle\psi_{j\downarrow}\psi_{j\uparrow}\rangle\approx\Delta$,
we can recast Hamiltonian (\ref{eq-h-site}) into a matrix form, i.e., the BdG Hamiltonian.
In particular, due to the presence of the staggered SO coupling,
we rewrite Hamiltonian (\ref{eq-h-site}) in terms of A-B sublattices by invoking the following representation:
$\psi_{2{\bf j}-\hat{\bf e}_y,s} \rightarrow a_{{\bf j}s}$ and
$\psi_{2{\bf j},s} \rightarrow b_{{\bf j}s}$.
After the transformation,
the odd-index (even-index) sites along the $y$ direction are mapped to those on A (B) sublattice, respectively.
$a$ and $b$ are the corresponding atomic operators of the sublattices.
As the results, the order parameter may also exhibit a staggered pattern, and we modify the uniform order parameter $\Delta$ by adding an additional term $\delta\Delta$: $-U\langle a_{{\bf j}\downarrow}^\dag a_{{\bf j}\uparrow}\rangle\equiv\Delta+\delta\Delta$ and $-U\langle b_{{\bf j}\downarrow}^\dag b_{{\bf j}\uparrow}\rangle\equiv\Delta-\delta\Delta$.
In the momentum ${\bf k}$ space, under the base $\Psi=(a_{{\bf k},\uparrow},b_{{\bf k},\uparrow},a_{{\bf k},\downarrow},b_{{\bf k},\downarrow},a_{-{\bf k},\downarrow}^\dag,b_{-{\bf k},\downarrow}^\dag,-a_{-{\bf k},\uparrow}^\dag,-b_{-{\bf k},\uparrow}^\dag)^T$,
the BdG Hamiltonian is expressed as
\begin{align}
	&H_{\rm BdG}({\bf{k}}) = \mathcal{H}_0({\bf{k}}) + \mathcal{H}_{\rm ext}({\bf{k}}) + \mathcal{H}_{\rm int}({\bf{k}})
	\,, \label{eq-hk-bdg}
\end{align}
with
\begin{align}
		\mathcal{H}_0({\bf{k}}) &= -[2t \cos (k_xd) +\mu]\tau_z + \delta\mu\tau_z\zeta_z \notag\\
		& -t[ \eta+ \cos(k_yd)] \tau_z\zeta_x - t \sin(k_yd) \tau_z\zeta_y \,,\label{eq-h-0-k}\\
		\mathcal{H}_{\rm ext}({\bf{k}}) &= 2 \alpha \sin(k_xd) \tau_z\sigma_y +\alpha[ \cos(k_yd) - \eta]\tau_z\sigma_x\zeta_y \notag\\
		& - \alpha \sin(k_yd) \tau_z\sigma_x\zeta_x + V_z \sigma_z \,,\\
		\mathcal{H}_{\rm int}({\bf{k}}) &= \Delta\tau_x + \delta\Delta \tau_x\zeta_z \,.\label{eq-h-int-k}
\end{align}
Here $\tau_{x,y,z}$ and $\zeta_{x,y,z}$ are Pauli matrices defined on the particle-hole and $a$-$b$ operator basis.

The ground state of the system is determined by the thermodynamic potential $\Omega$.
At zero temperature, it is given by
\begin{equation}
	\Omega= \frac{1}{4}\sum_{{\bf k},\nu} E_\nu({\bf k})\Theta[-E_{\nu}({\bf k})] + \mathcal{E}_0 \label{eq-thermo-pot}
\end{equation}
Here $E_\nu({\bf k})$ is the energy of the $\nu$th eigenstate of Hamiltonian (\ref{eq-hk-bdg}).
$\Theta(\cdot)$ is the Heaviside step function that is used to describe the Fermi-Dirac distribution at zero temperature.
The energy constant  $\mathcal{E}_0=(|\Delta|^2+|\delta\Delta|^2)/U
+\sum_{{\bf k}}[-2t\cos(k_xd)-\mu]$.
The order parameters $\Delta$ and $\delta\Delta$ can be obtained by self consistently solving the gap equations
\begin{equation}
	\frac{\partial \Omega}{\partial \Delta}=0 \quad \text{and}\quad
	\frac{\partial \Omega}{\partial \delta\Delta}=0 \,,
\end{equation}
while the filling factor $n$ can be obtained by the number equation
\begin{equation}
	n=-\frac{\partial \Omega}{\partial \mu} \,. \label{eq-number-eq}
\end{equation}
When $\Delta\neq0$ or $\delta\Delta\neq0$,
it outlines the superfluid phase region,
otherwise is a normal gas if they vanish.

\begin{figure*}[t]
	\centering
	\includegraphics[width=0.98\textwidth]{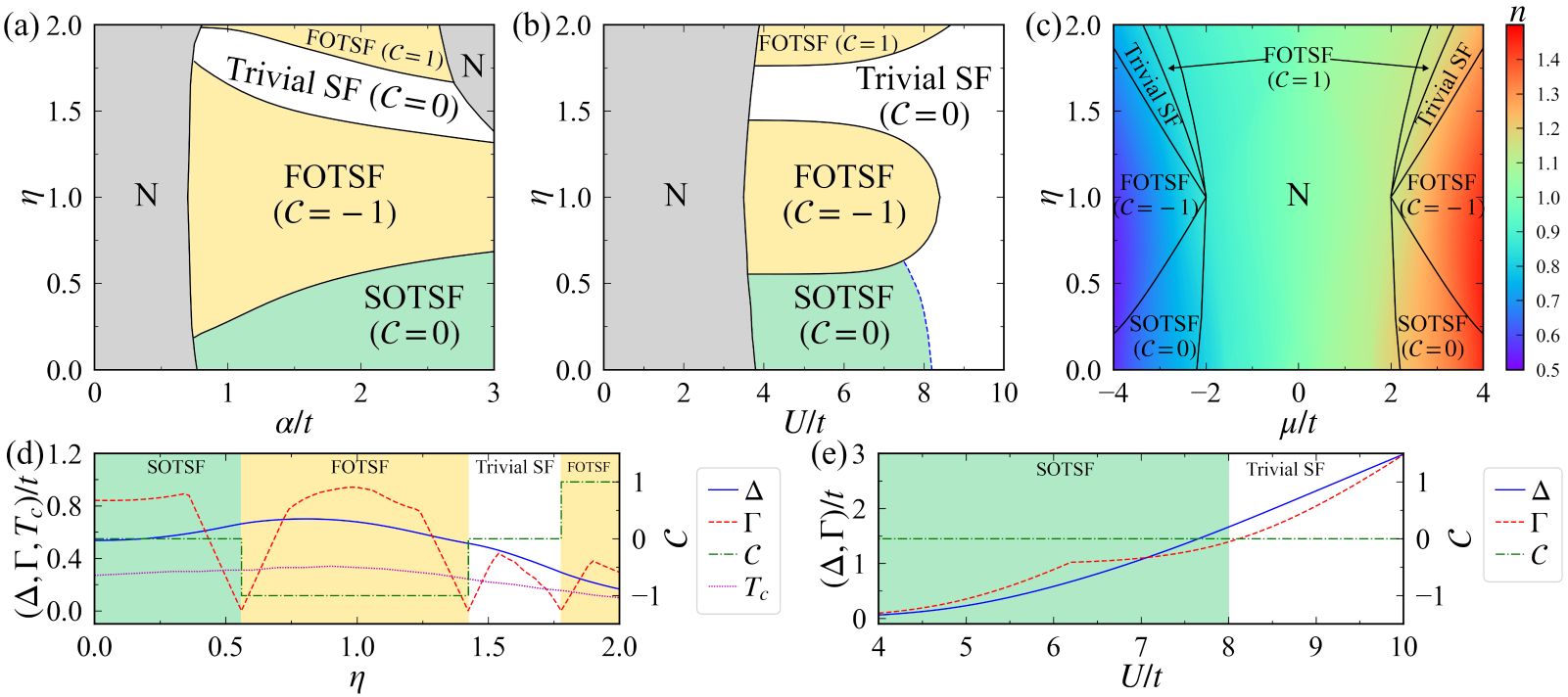}
	\caption{
		(a)-(c) Phase diagrams in the (a) $\alpha$-$\eta$, (b) $U$-$\eta$, and (c) $\mu$-$\eta$ planes.
		We set $(U,\mu)=(6t,3t)$ in (a), $(\alpha,\mu)=(2t,3t)$ in (b), $(\alpha,U)=(2t,6t)$ in (c),
		$V_z=2t$ and $\delta\mu=0$ in all panels.
		Trivial SF and N stand for the trivial superfluid and normal gas phase, respectively.
		The black solid (blue dashed) lines correspond to the phase region boundaries with (without) bulk-gap closing.
		The colors in (c) characterize the filling factor $n$.
		(d)-(e) The order parameter $\Delta$ (blue solid line), bulk gap $\Gamma$ (red dashed line), and Chern number $\mathcal{C}$ (green dash-dotted line) as functions of (d) $\eta$ and (e) $U$.
		We set $(\alpha,U,V_z)=(2t,6t,2t)$ in (d) and $(\alpha,\eta,V_z)=(2t,0.33,2t)$ in (e).
		In (d) we additionally show the critical temperature $T_c$ by the magenta dotted line.
	}
	\label{fig-phase}
\end{figure*}

\subsection{Phase diagrams}

The phase diagrams are displayed in FIG.\ref{fig-phase}(a) and (b) with different interaction strengths.
For simplicity, here we firstly consider the balanced A-B filling case $\delta\mu=0$.
Since our motivation for this work is based on the known FOTSF phase,
we directly tune $V_z$ to obtain it at $\eta=1$.
The superfluid phase region is composed of
the trivial superfluid, FOTSF, and SOTSF phases.
Moreover, SOTSF does not appear unless $\eta<1$,
yielding that $\eta$ is vital for processing the SOTSF transition.
The order parameters are found to be uniform in all the superfluid phases, i.e., $\delta\Delta=0$, due to the balanced A-B filling.
To distinguish the FOTSF and SOTSF phases, we inspect two topological invariants:
the Chern number $\mathcal{C}$ \cite{fukui2005chern}
and the quadrupole moment $q_{xy}$ \cite{benalcazar2017quantized}. The Chern number is widely applied in characterizing MEMs,
which is well known as bulk-edge correspondence.
It is defined by
\begin{equation}
	\mathcal{C}= \frac{1}{2\pi}\sum_{n}^\text{occ.}\int F_{nz} d{\bf k} \,,
\end{equation}
with Berry curvature defined as ${\bf F}_n=\nabla\times{\bf A}_n$ and Berry connection ${\bf A}_n=-i \langle u_{n{\bf k}}|\frac{\partial}{\partial{\bf k}}|u_{n{\bf k}} \rangle$.
The summation $\sum_{n}^\text{occ.}$ takes over all the occupied quasiparticle bands
and $|u_{n\bm{k}} \rangle$ is the $n$th occupied eigenstate.
The quadrupole moment can be calculated by means of the nested Wilson loops formalism \cite{franca2018anomalous}.
It refines the topologically protected charges that characterize the existence of MCMs.
In particular, the quadrupole moment is defined as \cite{benalcazar2017quantized}
\begin{equation}
q_{xy}=p_y^{\nu_x^+}p_x^{\nu_y^+}+p_y^{\nu_x^-}p_x^{\nu_y^-} \,,
\end{equation}
where the invariants $p_y^{\nu_x^\pm}$ and $p_x^{\nu_y^\pm}$ are calculated by
the Wannier sector polarizations (see Appendix \ref{sec-app-wilson}).
The system lies in SOTSF when $q_{xy}$ is half quantized,
otherwise is topological trivial when $q_{xy}=0$.

In FIG.\ref{fig-phase}(a),
we find two FOTSF regions with Chern numbers of opposite signs.
As mentioned in Sec.\ref{sec-model}, when $\eta=1$, Hamiltonian (\ref{eq-h-site}) has been known for supporting MEMs.
We find the desired SOTSF phase with $\mathcal{C}=0$ and $q_{xy}=1/2$ isolated from the trivial superfluid phase by the FOTSF regions.
In FIG.\ref{fig-phase}(b), we investigate the interaction effect on the phase transition.
The system transits to the superfluid phase as long as $U$ exceeds a threshold.
Besides the conventional FOTSF transition, we find the system undergoes a direct transition from the trivial superfluid phase to SOTSF, during which $\mathcal{C}$ remains zero but $q_{xy}$ changes instead.
The phase diagram reveals that the SOTSF transition is inadequate to be figured out by $\mathcal{C}$, instead its topological invariant is replaced by $q_{xy}$.

The influence of the chemical potential on the system is shown in FIG.\ref{fig-phase}(c).
Due to the particle-symmetry introduced by the order parameter,
the phase diagram is symmetric with respect to $\mu=0$ (i.e., half filling).
We find that the superfluid phase collapses in a wide range between $\mu\approx \pm2t$.
This is because, at the large Zeeman field $V_z$, the single-particle properties of the nearly half-filled system exhibit a band insulator \cite{Sun2013pra},
and the existence of the fermionic superfluidity is known to be manifestly suppressed \cite{Hu2006pra}.

We remark that the superfluid phases potentially exist even at finite temperature.
This can be qualitatively estimated by the critical temperature $T_c$ of the superfluid phases under the BdG approach
\footnote{At temperature $T$, the form of $\Omega$ in Eq.(\ref{eq-thermo-pot}) will be rewritten as $\Omega=\frac{k_BT}{4}\sum_{{\bf k},\nu} \ln f[-E_\nu({\bf k})] +\mathcal{E}_0$, where $f(E)=1/[\exp(E/k_BT)+1]$ is the Fermi-Dirac distribution. We assume the Boltzmann constant $k_B=1$ in the whole paper.},
as shown in FIG.\ref{fig-phase}(d).
In the tight-binding approximation, the hopping magnitude $t$ is typically of the order $0.1E_R$ \cite{Walters2013pra}.
Since the recoil energy $E_R\sim (\pi/3)^{2/3}E_F$  \cite{Hofstetter2002prl} ($E_F$ denoting the Fermi energy of the Fermi gas),
we can find $T_c$($\approx$$0.3t$) is of the order $0.1E_F$.
Hence the superfluid phases can be expected to survive at $T\sim 0.05E_F$ with current experimental techniques \cite{Greiner2005prl,Sobirey2021sci}.

\subsection{Topological features} \label{sec-topo-features}

\begin{figure*}[t]
	\centering
	\includegraphics[width=0.98\textwidth]{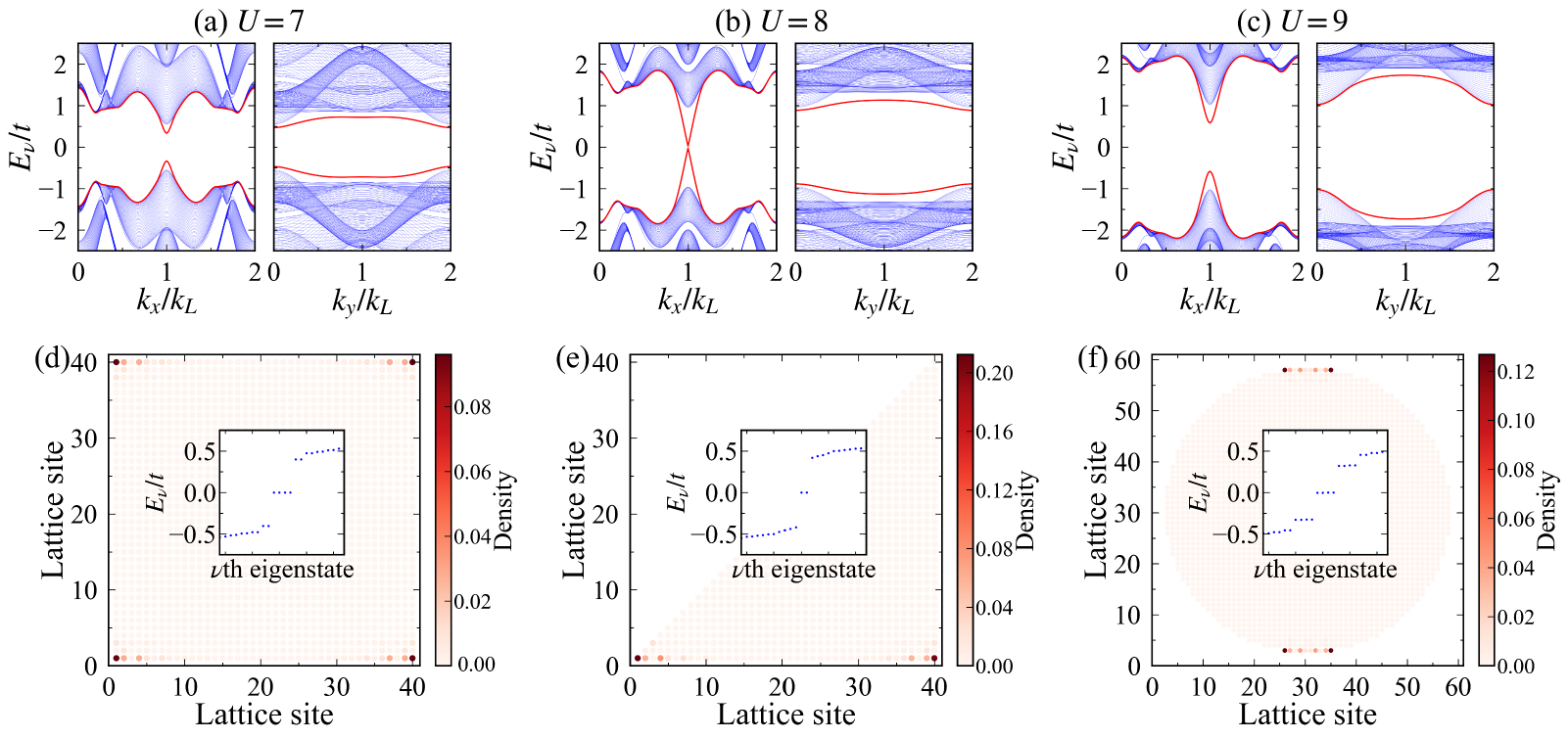}
	\caption{
		(a)-(c) The quasiparticle spectrum with cylindrical geometry for different $U$ at ($\alpha,\eta$)=($2t,0.33$).
		We set the open boundary condition in $x$ for left and $y$ for right of each panel.
		Other parameters are the same with FIG.\ref{fig-phase}(b).
		The twofold degenerate edge-state spectrum is highlighted by red solid lines.
		(d)-(f) The spatial distribution of the MCMs with the same parameters as (a) but in different boundary geometries.
		The colors characterize the number density occupied on each site.
		The inset shows the quasi-particle spectrum in the vicinity of zero energy.
		We employ the open boundary condition with different boundary geometries,
		while the Hamiltonian (\ref{eq-h-site}) remains the original square one.
		The edge size of the lattice sample is set as $L=40$ in (d) and (e), while the radius is set as 28 in (f).
	}
	\label{fig-corner-state}
\end{figure*}

The feature of the existed topological transitions relies on the
intrinsic bulk-boundary correspondence, but in various ways.
To extract the underlying properties of topological transition,
we plot the bulk gap $\Gamma$ and the Chern number $\mathcal{C}$ with respect to $\eta$.
In FIG.\ref{fig-phase}(d), by changing $\eta$,
the bulk gap closes and reopens at the transition between different topological phases.
This is consistent with the conventional physics picture,
in which the topological transition occurs in the company of the bulk gap closing as well as the changed Chern number.
The topological transition is known of the first order, and obeys the unambiguous bulk-edge correspondence.
The existing edge states are topologically protected in FOTSF unless the bulk gap closes.

However in FIG.\ref{fig-phase}(e), by changing $U$, both $\Delta$ and $\Gamma$ increase monotonically,
during which the unconventional topological transition without gap closing processes from the SOTSF to the trivial phase.
To capture the features of the SOTSF transition,
we use the cylindrical geometry by imposing a periodic boundary condition in the $x$ or $y$ direction.
The quasiparticle spectrum $E_\nu$ is obtained by diagonalizing the BdG Hamiltonian in spatial space,
shown in FIG.\ref{fig-corner-state}(a)-(c).
By increasing $U$, in the cylindrical geometry along the $y$ direction,
the edge-state gap closes at the critical point and reopens in the $x$ direction when system evolves into SOTSF.
In the cylindrical geometry along the $x$ direction, the edge-state gap remains open.
Therefore, the topological transition without bulk closing is of second order,
as signaled by the edge-state gap.
The second-order topological transition follows the edge-corner correspondence \cite{ezawa2020edgecorner} rather than the bulk-edge correspondence.
In other words, the corners of the lattice plaquette play the role of boundaries while the edges are regarded as the bulk.
For this sake, the SOTSF phase is also known as being boundary obstructed \cite{tiwari2020chiral}.

The Hamiltonian (\ref{eq-hk-bdg}) of the system preserves particle-hole symmetry:
$\mathcal{P}H_{\rm BdG}({\bf k})\mathcal{P}^{-1}=-H_{\rm BdG}(-{\bf k})$ with $\mathcal{P}=\tau_y\sigma_y\mathcal{K}$
and $\mathcal{K}$ standing for the complex conjugation.
In the presence of $V_z$, the time-reversal symmetry is broken.
Meanwhile, the spatial-inversion symmetry is also satisfied: $\mathcal{I}H_{\rm BdG}({\bf k})\mathcal{I}^{-1}=H_{\rm BdG}(-{\bf k})$ where $\mathcal{I}=\sigma_z\zeta_x$ anticommutes with $\mathcal{P}$.
Therefore, the system belongs to the D class \cite{Geier2018prb,Khalaf2018prb}.
Due to the bulk-edge correspondence of SOTSF,
we can see that the MCMs are fourfold degenerate in the square lattice due to the two symmetries.
This is shown in FIG.\ref{fig-corner-state}(d) for the square boundary geometry and (f) for the circular one.
However, the degeneracy of MCMs is affected by the adjustable boundary geometry,
and reduces to be twofold in a triangular lattice (FIG.\ref{fig-corner-state}(e)).
This is a ubiquitous feature of the SOTSF phase \cite{yan2018majorana,ezawa2018topological,wu2019plane}.

\subsection{Imbalanced A-B filling}

\begin{figure}[t]
	\centering
	\includegraphics[width=0.49\textwidth]{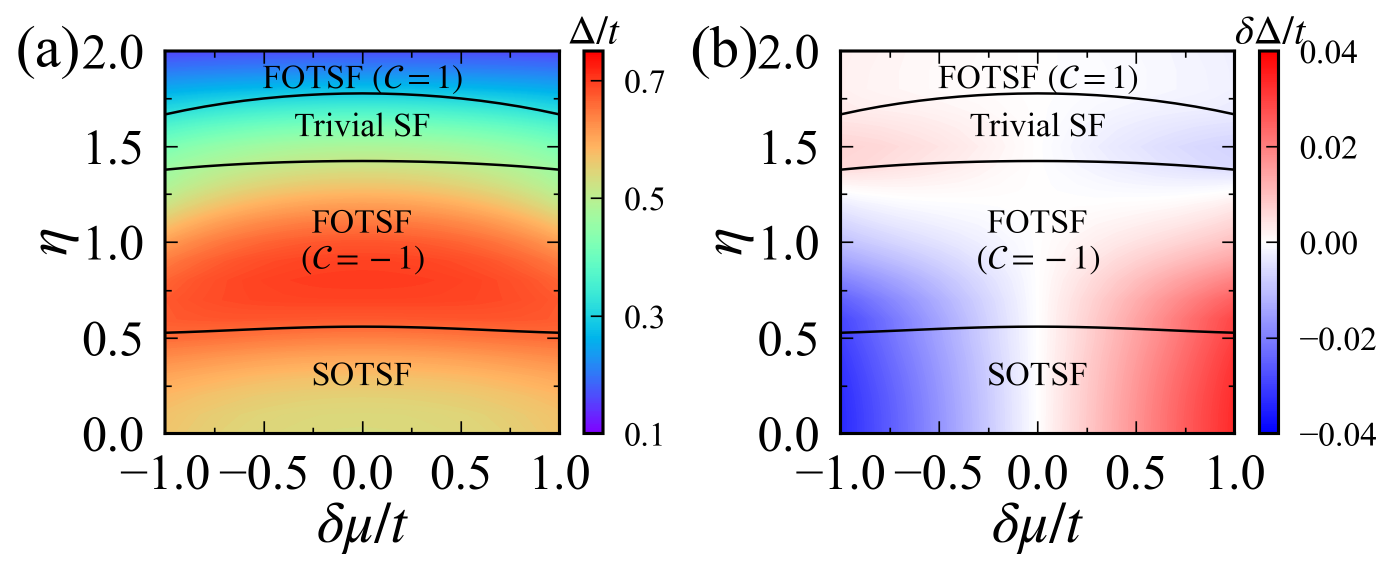}
	\caption{
		The order parameters (a) $\Delta$ and (b) $\delta\Delta$ in the parameter plane of $\delta\mu$ and $\eta$.
		The colors characterize the magnitudes of order parameters,
		and specifically in (b), the white region indicates the uniform order parameter with $\delta\Delta=0$.
		The black-solid lines outline the boundaries of the phase regions.
		We set ($\alpha,U,\mu,V_z$)=($2t,6t,3t,2t$).
	}
	\label{fig-dmu}
\end{figure}

We now investigate the influence of the imbalanced A-B filling case, i.e., $\delta\mu\neq0$.
The order parameters and the associated phase diagram in the $\delta\mu$-$\eta$ plane are shown in FIG.\ref{fig-dmu}.
We can find that the modulated order parameter $\delta\Delta$ changes monotonically with respect to $\delta\mu$ and remains zero when $\delta\mu=0$.
Furthermore, the magnitude of $\delta\Delta$ is obviously less than $\Delta$,
revealing a weak staggered pattern.
As the results, the phase boundaries between different superfluid phases varies nearly independently of $\delta\mu$.

The presence of the modulated density triggered by $\delta\mu$ does not affect the topological features of the superfluids.
This is because both $\delta\mu\tau_z\zeta_z$ in Eq.(\ref{eq-h-0-k}) and $\delta\Delta\tau_x\zeta_z$ in Eq.(\ref{eq-h-int-k}) preserve the particle-hole and inversion symmetries discussed in Sec.\ref{sec-topo-features}.
Therefore, the topological phases in the phase diagrams are robust against the modulated density and order parameter.

\section{Discussions} \label{sec-dis}

In practice, the model Hamiltonian (\ref{eq-h-site}) can be engineered as follows.
A direct way for generating the SO coupling is based on optical fields with standing-wave modes
like $M({\bf r})\sim\sin(k_Lx)\cos(k_Ly)+i\cos(k_Lx)\sin(k_Ly)$ \cite{Liu2014soc}.
In this setup, the onsite spin hybridization is thus dramatically suppressed and ignorable because of the odd parity of $M({\bf r})$ with respect to each site center,
by contrast the NN coupling is dominant and generates the SO coupling.
The parameter that characterizes the staggered pattern is thereby obtained by
$\eta' = \mathcal{I}_1/\mathcal{I}_2$,
where $\mathcal{I}_n = \int d{\bf r}\cos(k_Lx)\sin(k_Ly)W^*({\bf r}+n\hat{\bf e}_y)W[{\bf r}+(n+1)\hat{\bf e}_y]$
and $W({\bf r})$ stands for the localized Wannier wave function on each site.
The alternative way for generating the SO coupling is assisted by an additional gradient field with the strength $\delta$ in both two directions.
The original NN hopping will be prohibited due to the gradient field.
Motivated by the laser-assisted tunneling technique \cite{Aidelsburger2013prl,Miyake2013prl},
by using Raman optical fields whose frequency offset matches $\delta$,
not only the SO coupling can be engineered, but also the NN hopping is restored.
In this setup, the magnitudes of the SO coupling and NN hopping can be separately tunable.
This is attainable by using two groups of optical fields that separately drives different transitions due to the selection rule.

In order to observe the MCMs localized in lattice corners,
it is intuitive to produce the real-space sharp edges by preparing the external confinement \cite{Goldman2013pnas}.
We remark that since $\eta$ is responsible for multi-order topological transitions,
it provides an alternative way for creating the artificial topological interfaces \cite{Reichl2014pra,Goldman2016pra,Leder2016ncomms,Meier2016ncomms,An2017science,Irsigler2019prl} by preparing the spatially-dependent $\eta$.
The MCMs inside the band gap may be signaled via the Bragg spectra \cite{Goldman2012prl}
or visualized from the Wannier-Stark spectrum \cite{Poddubny2019prb}.

In summary, we demonstrate an experimental feasible scheme for engineering SOTSF that supports MCMs.
The double-well superlattice potential imposes a staggered pattern to the NN hopping and SO coupling,
which leads to multi-order topological phase transitions.
The FOTSF and SOTSF transitions are separately characterized by different topological invariants as well as band gap signals, revealing various bulk-boundary correspondences.
The proposal is feasible by means of current experimental techniques, and can pave the way for exploring SOTSF and the associated MCMs.

\begin{acknowledgments}
	This work is supported by
	National Natural Science Foundation of China (Grants No. 11474271, 11674305, and 11704367).
\end{acknowledgments}

\appendix

\section{Nested Wilson loops} \label{sec-app-wilson}

\begin{figure}[t]
	\centering
	\includegraphics[width=0.48\textwidth]{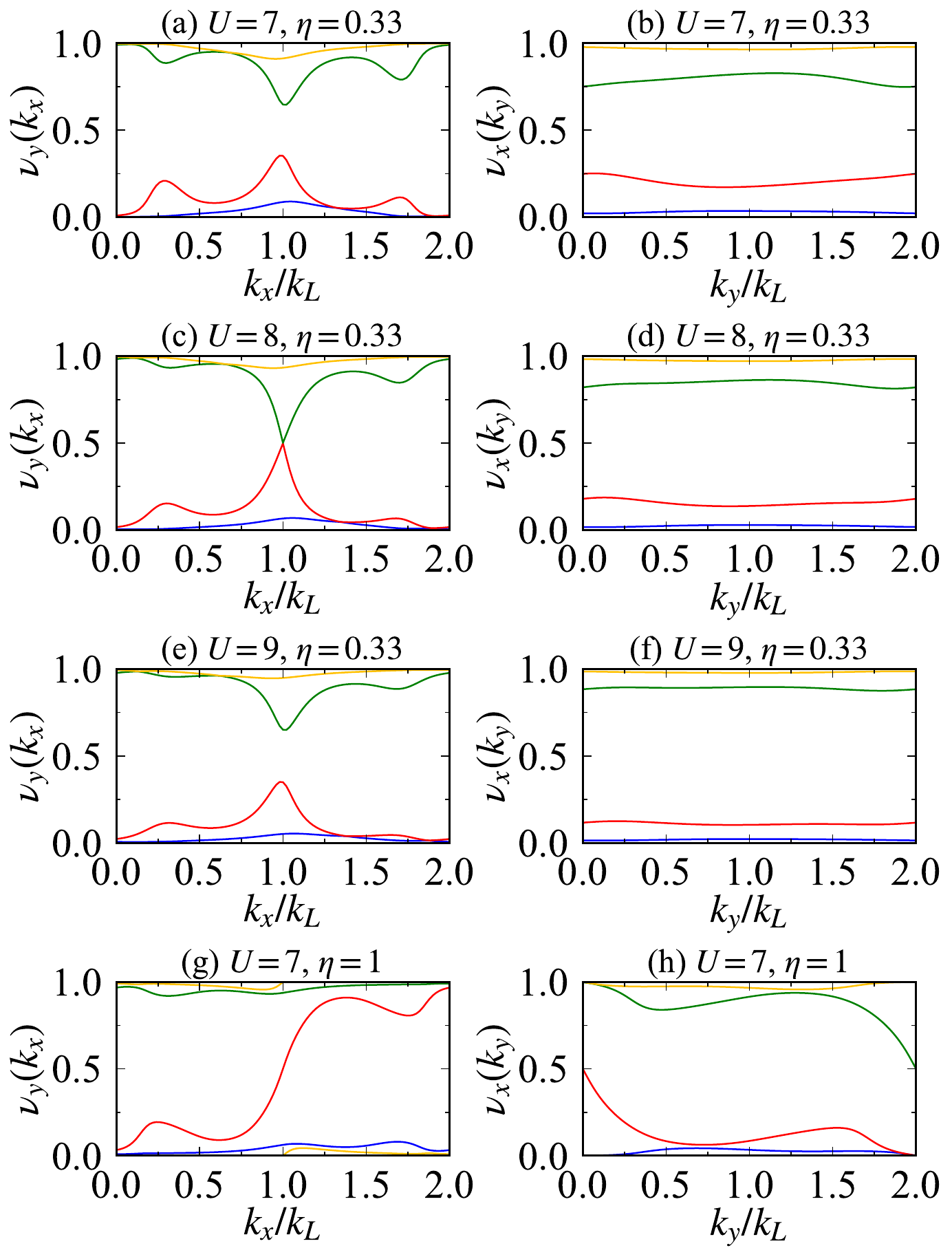}
	\caption{Band structure of the Wannier sectors with open boundaries in a single direction:
	$y$ (left column) or $x$ (right column).
	The colors stand for different sectors.
	The band gap closes at (c) $k_x=k_L$, (g) $k_x=k_L$ and (h) $k_y=0$ ($2k_L$).
	(a) and (b) are plotted at ($U$,$\eta$)=($7t$,0.33), corresponding to the trivial superfluid.
	(c) and (d) are plotted at ($U$,$\eta$)=($8t$,0.33), corresponding to the gap-closing point (phase transition).
	(e) and (f) are plotted at ($U$,$\eta$)=($9t$,0.33), corresponding to SOTSF.
	(g) and (h) are plotted at ($U$,$\eta$)=($7t$,1), corresponding to FOTSF.
	Other parameters are the same as in FIG.\ref{fig-phase}(b).}
	\label{fig-wannier}
\end{figure}

Here we present the technical details for calculating the quadrupole momentum by means of nested Wilson loops.
We first diagonalize the total eight-band Hamiltonian and obtain the $N_\text{occ}=4$ occupied states $|u_{\bf k}^m\rangle$.
Then we define the Wilson loop operators $\mathcal{W}_{x,{\bf k}}$ and $\mathcal{W}_{y,{\bf k}}$ in the $x$ and $y$ directions, where ${\bf k}=(k_x,k_y)$ is the initial point of the Wilson loop operator.
For the $L\times L$ square lattice, by using the obtained occupied states, we define $N_\text{occ}\times N_\text{occ}$-dimensional matrix $[G_{x,{\bf k}}]^{mn}=\langle u^m_{{\bf k}+\Delta {\bf k}_x} | u^n_{{\bf k}}\rangle$, where $\Delta {\bf k}_{x,y} = \frac{2\pi}{L}\hat{\bf e}_{x,y}$.
Thus we can define the Wilson loop operator in the discrete limit
\begin{equation}
W_{x,{\bf k}} = G_{x,{\bf k}+L\Delta {\bf k}_x}\dots G_{x,{\bf k}+\Delta {\bf k}_x}G_{x,{\bf k}} \,.
\end{equation}
Because of the discretization of momentum ${\bf k}$, the matrix $G$ is not a unitary matrix.
It can be mapped to be unitary by the singular value decomposition (SVD) at each discretized momentum: $G_{x,{\bf k}}=UDV^\dag$.
In this way, by redefining $F_{x,{\bf k}}=UV^\dag$, we can rewrite the unitary Wilson loop operator as
\begin{equation}
\mathcal{W}_{x,{\bf k}} = F_{x,{\bf k}+L\Delta {\bf k}_x}\dots F_{x,{\bf k}+\Delta {\bf k}_x}F_{x,{\bf k}} \,.
\end{equation}
Under the periodic boundary condition of a torus geometry, the eigenequation of the Wilson loop operator is expressed as
\begin{equation}
\mathcal{W}_{x,{\bf k}}  |\nu_{x,{\bf k}}^{\pm,r} \rangle = e^{2\pi i \nu_x^j(k_y)} | \nu_{x,{\bf k}}^{\pm,r} \rangle \,.
\end{equation}
Here $\pm$ represent two different Wannier sectors, and $r=1\dots N_{occ}/2$.
The nested Wilson loops along the $y$ direction is thus obtained as
\begin{equation}
\tilde{\mathcal{W}}_{y,{\bf k}_x}^{\pm}=F_{y,{\bf k}+L\Delta {\bf k}_y}^{\pm}\dots F_{y,{\bf k}+\Delta {\bf k}_y}^{\pm}F_{y,{\bf k}}^{\pm} \,,
\end{equation}
where
\begin{equation}
F^{\pm}_{y,{\bf k}}=\langle w_{x,{\bf k}+\Delta {\bf k}_y}^{\pm,r}|w_{x,{\bf k}}^{\pm,r}\rangle \,,\,
|w_{x,{\bf k}}^{\pm,r}\rangle = \sum_{n =1}^{N_{occ}} |u_{{\bf k}}^n\rangle [\nu_{x,{\bf k}}^{\pm,r}]^n \,.
\end{equation}
$[\nu_{x,{\bf k}}^{\pm,r}]^n$ are the components of the Wilson loop eigenstate $|\nu_{x,{\bf k}}^{\pm,r}\rangle$.
The Wannier sector polarization is given by
\begin{equation}
p_y^{\nu_x^{\pm}}=-\frac{i}{2\pi}\frac{1}{L}\sum_{k_x}\mathrm{Log} [\tilde{\mathcal{W}}_{y,{\bf k}_x}^{\pm}] \,,
\end{equation}
It is specified as a $\mathbb{Z}_2$ topological index: $p_y^{\nu_x^{\pm}}\in \{0,1/2\}$.
Likewise, the other topological index $p_x^{\nu_y^{\pm}}$ can be obtained following the same approach.

The quadrupole invariant $q_{xy}$ is expressed as
\begin{equation}
q_{xy}=p_y^{\nu_x^+}p_x^{\nu_y^+}+p_y^{\nu_x^-}p_x^{\nu_y^-} \,.
\end{equation}
In FIG.\ref{fig-wannier} we display the band structure of the Wannier sectors in a cylindrical geometry.
In the SOTSF phase with $(p_x^{\nu_y^{\pm}},p_y^{\nu_x^{\pm}})=(1/2,1/2)$, we obtain a half quantized quadrupole moment $q_{xy}$.
However in the trivial superfluid phase, we obtain $(p_x^{\nu_y^{\pm}},p_y^{\nu_x^{\pm}})=(0,1/2)$, $(1/2,0)$, or $(0,0)$,
resulting in $q_{xy}=0$.
Therefore, $q_{xy}$ can play the role of a topological invariant featuring in the transition between the trivial superfluid and SOTSF phases.

We note that $q_{xy}$ is ill defined and hence not available for the FOTSF phase as well as the gap-closing point.
This is because the band structure of Wannier sectors is gapless over the entire Brillouin zone,
as shown in FIG.\ref{fig-wannier}(c), (g) and (h).
But the transition from FOTSF to other phases can still be characterized by the Chern number as in the conventional physics picture.

\vfill
\bibliographystyle{apsrev4-1}
\bibliography{bib}
\end{document}